# High frequency permeability of magnonic metamaterials with magnetic inclusions of complex shape

O. Dmytriiev[1,6,a], M. Dvornik[1], R. V. Mikhaylovskiy[1], M. Franchin[2], H. Fangohr[2], L. Giovannini[3], F. Montoncello[3], D. V. Berkov[4], E. K. Semenova[4], N. L. Gorn[4], A. Prabhakar[5], and V. V. Kruglyak[1,b]

1. School of Physics, University of Exeter, Exeter, EX4 4QL, United Kingdom

2. School of Engineering Sciences, University of Southampton, SO17 1BJ, United Kingdom

3. Dipartimento di Fisica, Università di Ferrara, Via G. Saragat 1, 44122 Ferrara, Italy

4. Innovent Technology Development, Pruessingstrasse, 27B, D-07745, Jena, Germany

5. Indian Institute of Technology Madras, Chennai, 600036, India

6. Institute of Magnetism, Kiev, 03142, Ukraine

**Abstract**

We present a method of calculation of the effective magnetic permeability of magnonic metamaterials containing periodically arranged magnetic inclusions of arbitrary shapes. The spectrum of spin wave modes confined in the inclusions is fully taken into account. Within the scope of the proposed method, we compare two approaches. The first approach is based on a simple semi-analytical theory that uses the numerically calculated susceptibility tensor of an isolated inclusion as input data. Within the second approach, micromagnetic packages with periodic boundary conditions (PBC) are used to calculate the susceptibility of a single 2D periodic array of such inclusions, with the whole 3D metamaterial consisting of a stack of such arrays. To calculate the susceptibility tensor of an isolated inclusion, we have implemented and compared two different methods: (a) a micromagnetic method, in which we have employed three different micromagnetic packages: the finite element package NMAG and the two finite differences packages OOMMF and MicroMagus; and (b) the modified dynamical matrix method. To illustrate the methodology, we have calculated the effective permeability of a metamaterial consisting of a stack of hexagonal arrays of magnetic nanodisks in a non-magnetic matrix. The range of geometrical parameters for which such a metamaterial is characterized by the negative permeability has been identified. The critical comparison of the different micromagnetic packages and the dynamical matrix method (based on the calculation of the susceptibility tensor of an isolated inclusion) has demonstrated that their results agree to within 3 %.

---

[a] Also for correspondence: O.Dmytriiev@exeter.ac.uk

[b] Corresponding author: V.V.Kruglyak@exeter.ac.uk

# I. INTRODUCTION

The recent progress in electromagnetic metamaterials has been fueled by the discovered ability to design their unusual properties[1,2] via tweaking the geometry and structure of the constituent "meta-atoms"[3]. Along with negative permittivity, negative permeability is one of the necessary features for the design of negative refractive index metamaterials. A metamaterial designer can achieve negative permeability via geometrical control of high frequency currents, e.g. in arrays of split ring resonators[4], or alternatively can rely on spin resonances in natural magnetic materials[5,6], as was suggested by Veselago in Ref. 1. However, the age of nanotechnology sets an intriguing quest for additional benefits to be gained by nano-structuring natural magnetic materials into so called *magnonic* metamaterials, in which the frequency and strength of resonances based on spin waves (magnons)[7] are determined by the geometry and magnetization configuration of meta-atoms. Spin waves can have frequencies up to hundreds of GHz (in the exchange dominated regime)[6-9] and have already been shown to play an important role in the high frequency magnetic response of composites containing magnetic inclusions of cylindrical[10-12] and spherical[13-17] shape.

The majority of analytical models of the effective permeability of magnetic composites and metamaterials employ the macrospin approximation, in which each magnetic inclusion within a non-magnetic matrix is considered as a single giant spin and is therefore characterized by a single magnetic resonance. However, it is well known that the spin wave spectrum of magnetic nano-structures and nano-elements has a complex structure, featuring series of resonances due to spatially non-uniform spin wave modes[18-22]. Each of the resonances is expected to contribute to the susceptibility tensor of the magnetic constituents and correspondingly to the permeability tensor of the whole metamaterial. The resonance frequencies can be controlled and reconfigured by the external magnetic[19-24] and electric[25,26] fields, and the same functionalities could therefore be inherited by the magnonic metamaterials.

In this paper, we demonstrate a method of calculation of the effective permeability that takes full account of the complex spectrum of the metamaterial's individual magnetic constituents. In this method, the susceptibility tensor of an isolated inclusion is calculated numerically and then used as an input to an analytical expression for the permeability of the whole metamaterial. To find the susceptibility tensor of the isolated inclusion, different approaches have been used. In one of them, we have performed full-scale numerical micromagnetic simulations using three different micromagnetic packages: a finite element based package NMAG[27] and two finite difference based packages OOMMF[28] and MicroMagus[29]. In the other approach, the dynamical matrix method, in which the system of linearized equations of motion of magnetic moments is solved to find the normal modes of a system[30], has been modified to facilitate the susceptibility calculations. The methods have been applied to a model metamaterial representing an array of magnetic nano-disks embedded into a non-magnetic matrix. In particular, we have been able to determine the region of geometrical parameters of such a metamaterial, in which one of the components of the permeability tensor becomes negative within a certain frequency range. The predictions of the method are compared with calculations based on micromagnetic simulations with the use of periodic boundary conditions (PBCs) and also with macrospin calculations. Furthermore, we use the calculations to compare the different micromagnetic methods in order to evaluate the accuracy to be expected from micromagnetic simulations.

In principle, the proposed method could be considered as an extension of the concept from Ref. 6 of using magnonic resonances to tailor effective permeability of metamaterials. However, we note that the stack of thin films studied in Ref. 6 is treatable analytically. In practice however, one might either want or have to deal with alternative realizations of the concept, i.e. to use magnonic "meta-atoms" of a different shape. For example, this could be dictated either by limitations of the available nanofabrication tools or by needs for permeability with a specific frequency dependence. Such more complex magnonic metamaterials would not necessarily allow a simple analytical treatment while numerical simulations of extended samples might present too high demands on computational resources. The proposed method circumvents the problem, to some degree in the spirit of approaches developed in Refs. 31,32.

## II. PERMEABILITY OF A MAGNONIC METAMATERIAL

### A. Analytical model

Let us consider an idealized case of an infinitely extended 3D metamaterial. To enable a meaningful introduction of the effective permeability $\hat{\mu}$, the wavelength of electromagnetic waves should be much greater than the characteristic dimensions of the magnetic inclusions and the lattice constant of the metamaterial. Then, we can use the standard "macroscopic" relation between the high frequency magnetic induction (**B**) and magnetic field (**H**) of the electromagnetic wave:

$$\mathbf{B} = \mathbf{H} + 4\pi\mathbf{M} = \hat{\mu}\mathbf{H}, \tag{1}$$

where **M** is the dynamic part of the magnetization that is spatially averaged of the volume of the metamaterial ("macroscopic magnetization"), and the permeability is defined in the frequency domain, $\hat{\mu} = \hat{\mu}(\omega)$. We generally denote macroscopic quantities by capital letters and microscopic ones by lower case ones. In particular, the static spatially averaged macroscopic magnetization is denoted as $\mathbf{M}_0$. Besides dynamic magnetic field **H**, there is also external spatially uniform static magnetic field $\mathbf{H}_{\text{bias}}$.

Permeability $\hat{\mu}$ is related to macroscopic susceptibility of the whole metamaterial $\hat{\chi}$ as

$$\hat{\mu} = \hat{I} + 4\pi\hat{\chi}. \tag{2}$$

Susceptibility $\hat{\chi}$ can be found by calculating the response of the volume averaged magnetization of the metamaterial to an external ac uniform magnetic field, which in general is a complicated problem to compute. The problem is simplified by assuming that the metamaterial represents a periodic lattice of magnetic elements ('inclusions') that are identical in terms of both their shape and material properties. Then, permeability $\hat{\mu}$ of the metamaterial can be related to the susceptibility a single inclusion ($\hat{\chi}_{\text{incl}}$) via a simple equation. The problem of finding the susceptibility a single inclusion is significantly simpler than that of finding the susceptibility of the whole metamaterial. The two susceptibilities differ due to the dipole-dipole interaction between inclusions inside the metamaterial. If this interaction is absent, $\hat{\chi}$ and $\hat{\chi}_{\text{incl}}$ differ only by a factor equal to volume fraction ("filling factor") of the magnetic inclusions in the metamaterial $\rho$ ($0 < \rho < 1$). Indeed, macroscopic magnetization **M** can be written as

$$\mathbf{M} = \rho \overline{\mathbf{m}}, \qquad (3)$$

where $\overline{\mathbf{m}}$ is the dynamic "microscopic magnetization" (i.e. one obtained via spatial averaging over the volume of a single magnetic inclusion), with the corresponding static magnetization denoted as $\overline{\mathbf{m}}_0$:

$$\overline{\mathbf{m}} = \frac{1}{V} \iiint_V \mathbf{m} \, dV, \qquad (4)$$

where $V$ is the volume of the magnetic inclusion. The dynamic magnetization $\overline{\mathbf{m}}$ is related to the local microscopic dynamic magnetic field $\mathbf{h}'$ through susceptibility tensor $\hat{\chi}_{incl}$, which is defined in the frequency domain as

$$\overline{\mathbf{m}} = \hat{\chi}_{incl} \mathbf{h}'. \qquad (5)$$

Each magnetic inclusion by itself is characterized by a complex spectrum of spatially nonuniform spin wave modes[18-22]. Due to confinement effects, these modes can generally couple to the electromagnetic field, even if it can be considered as uniform on the scale of the inclusion's dimensions. Hence, each resonant mode contributes to tensor $\hat{\chi}_{incl}$, thereby making its frequency dependence very intricate. However, micromagnetic simulations allow one to calculate $\hat{\chi}_{incl}$ of realistic inclusions of arbitrary shapes.

The macroscopic and local microscopic dynamic magnetic fields differ by the local dynamic dipolar field $\mathbf{h}_{dyn}$ created by the dynamic magnetization of magnetic inclusions:

$$\mathbf{h}' = \mathbf{H} + \mathbf{h}_{dyn}. \qquad (6)$$

Besides dynamic dipolar field $\mathbf{h}_{dyn}$, there is static magnetic field $\mathbf{h}_{stat}$, which is created by the static magnetization of the same inclusions. Similar to the dynamic part, local microscopic static magnetic field $\mathbf{h}'_0$ is given by

$$\mathbf{h}'_0 = \mathbf{H}_{bias} + \mathbf{h}_{stat} \qquad (7)$$

In the present case of the electromagnetic wavelength that is much greater than the distance between inclusions, we can express the local dipolar field via tensor $\hat{N}$ that is analogous to the tensor of demagnetizing coefficients:

$$\mathbf{h}_{dyn} = -4\pi \hat{N} \overline{\mathbf{m}}. \qquad (8)$$

Substituting equations (2)-(8) into equation (1), we obtain the following general expression for effective permeability $\hat{\mu}$:

$$\hat{\mu} = \hat{I} + 4\pi\rho \hat{\chi}_{incl} \left( \hat{I} + 4\pi \cdot \hat{N} \hat{\chi}_{incl} \right)^{-1}. \qquad (9)$$

To illustrate the method outlined above, we evaluate the effective permeability of a metamaterial that consists of magnetic disks placed in nodes of a hexagonal lattice (Fig. 1). The distance between layers is taken to be much larger than the edge-to-edge separation between disks within each layer. Then, one can neglect the magneto-dipolar interaction between layers, and the metamaterial can therefore be considered as a stack of quasi-two-dimensional planes.

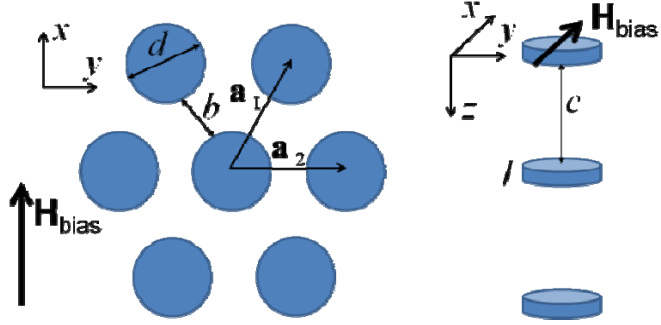

**Fig. 1.** (Color online) The geometry of the metamaterial consisting of magnetic discs in a non-magnetic matrix is shown. The discs are located in nodes of a hexagonal lattice. The disk diameter is $d = 195$ nm, the in-plane edge-to-edge separation is $b = 20$ nm, the distance between the layers is $c = 140$ nm and is much greater than disk thickness $l = 5$ nm. The filling factor is $\rho = 2.48\%$.

In the general case, dynamic magnetodipolar field $\mathbf{h}_{dyn}$ inside the given inclusion depends on the spatial distributions of the dynamic magnetization (mode profiles) inside all other inclusions. We can obtain $\mathbf{h}_{dyn}$ using the multipole expansion method, i.e. by expanding the magneto-dipolar field created by each inclusion over the multipole moments of its dynamic magnetization configuration. For simplicity, in the present paper, we restrict ourselves only to the first (dipolar) term in this expansion. Then, $\mathbf{h}_{dyn}$ can be expressed in terms of averaged magnetizations $\overline{\mathbf{m}}$ only as

$$\mathbf{h}_{dyn} = \sum_{\mathbf{R} \neq 0} \frac{3\mathbf{R}(\mathbf{p}_m \mathbf{R}) - \mathbf{p}_m \mathbf{R}^2}{\mathbf{R}^5}, \tag{10}$$

where $\mathbf{p}_m = \overline{\mathbf{m}} V$ is the total dynamic magnetic moment of an inclusion and the summation is performed over all nodes of one hexagonal layer. The sum in equation (10) converges since we perform only a 2D summation.

In the framework of the dipolar approximation, the lattice of magnetic disks is effectively replaced by the lattice of magnetic moments $\mathbf{p}_m$ located in the centers of the disks. However, in contrast to the macrospin approximation, each magnetic moment $\mathbf{p}_m$ here inherits the complex excitation spectrum of a single magnetic disk, both in terms of the frequencies and strength of the resonances. This approximation is valid when the energy of the interaction between disks is small enough (so that it can be considered as perturbation) and does not lead to the modification of the inclusion's ground state (i.e. for large distances between the disks). However, as the distance between the disks decreases, the profiles of some modes could be modified by the interaction, leading to significant changes in both the dynamic dipolar moments and frequencies of the modes, as observed in Ref. 33. This case requires a special consideration and shall therefore be considered in future publications.

In principle, magnetic moment $\mathbf{p}_m$ in equation (9) could vary from one inclusion to another within collective magnonic modes of the metamaterial with a nonzero wavevector[21]. However, in the case of an infinitely extended metamaterial considered here, only modes with the spatial dependence following that of the incident electromagnetic wave could be excited. The frequencies of the dominant modes for a single disk are in the range of tens of GHz[18,19], which corresponds to electromagnetic waves in the centimeter wavelength range. The associated spatial variation of the phase of collective magnonic modes is extremely slow and therefore can be neglected. Hence, we limit the consideration to spatially uniform excitations of the metamaterial. Then, vector $\mathbf{p}_m$ in equation (10) can be taken out of the sum, and, according to definition (8), we can calculate tensor $\hat{N}$ as

$$N_{ij} = -\frac{V}{4\pi} \sum_{R \neq 0} \frac{3R_i R_j - \delta_{ij} R^2}{R^5}. \tag{11}$$

In contrast to the tensor of demagnetizing coefficients, whose trace is equal to unity, the trace of tensor $\hat{N}$ here is equal to zero, since we do not take into account the magnetic moment in the point in-which we calculate the magnetic field. Instead, magnetic interactions within each magnetic inclusion are taken into account in the full micromagnetic simulations described below. To calculate tensor $\hat{N}$, we note that any lattice site of a 2D hexagonal lattice is determined by vector

$$\mathbf{R} = n_1 \mathbf{a}_1 + n_2 \mathbf{a}_2 = a\frac{\sqrt{3}}{2} n_2 \hat{\mathbf{e}}_x + a\left(n_1 + \frac{1}{2}n_2\right)\hat{\mathbf{e}}_y,$$

where $a$ is the lattice period (center-to-center separation between the disks). Substituting this into equation (11), we calculate tensor $\hat{N}$ numerically to yield

$$\hat{N} = \begin{pmatrix} -\xi & 0 & 0 \\ 0 & -\xi & 0 \\ 0 & 0 & 2\xi \end{pmatrix}, \tag{12}$$

where $\xi = \dfrac{0.44V}{a^3}$ and we did the summation for $n_1, n_2$ in the finite large interval from -1000 to 1000 what leads to the accuracy $\pm 0.0001$ (this accuracy much higher than the rounding of $\xi$ done by us)

In the numerical calculation of susceptibility tensor of an isolated disk $\hat{\chi}_{\text{incl}}$, one should take into account that, according to equation (7), the local static magnetic field, experienced by each magnetic inclusion within the metamaterial and which must therefore be used in the simulations, differs from external field $\mathbf{H}_{\text{bias}}$ applied to the metamaterial. The difference is equal to local static dipolar field $\mathbf{h}_{\text{stat}}$ originating from the other magnetic inclusions. This means that, if we simulate the susceptibility of the isolated disk for the external field $\mathbf{h'}_0$, the result obtained for the permeability of the metamaterial is valid for the applied field of

$$\mathbf{H}_{bias} = \mathbf{h'}_0 - \mathbf{h}_{stat}.  \qquad (13)$$

The relation between the static interaction field and the static magnetization is the same as for their dynamical counterparts, so that we can follow equation (8) to write

$$\mathbf{h}_{stat} = -4\pi \hat{N} \overline{\mathbf{m}}_0. \qquad (14)$$

### B. Macrospin approximation

In this subsection, we use the macrospin approximation to derive a simple analytical expression for susceptibility tensor $\hat{\chi}_{incl}$ and corresponding permeability tensor $\hat{\mu}$. This result is subsequently compared with susceptibility tensor $\hat{\chi}_{incl}$ calculated numerically. In the macrospin approximation and neglecting damping, the expression for tensor $\hat{\chi}_{incl}$ follows directly from the solution of the Landau-Lifshits equation

$$\frac{\partial \overline{\mathbf{m}}}{\partial t} = \gamma [\mathbf{h}_{eff} \times \overline{\mathbf{m}}]$$

for an isolated inclusion (a nanodisk in our case) with the effective field

$$\mathbf{h}_{eff} = \mathbf{H}_{bias} - 4\pi \hat{N}_{disk} \overline{\mathbf{m}} + \mathbf{h'}(t)$$

in terms of $\mathbf{h'}(t)$, where

$$\hat{N}_{disk} = \begin{pmatrix} n_\parallel & 0 & 0 \\ 0 & n_\parallel & 0 \\ 0 & 0 & n_\perp \end{pmatrix}$$

is the tensor of demagnetizing coefficients for an isolated disk. For thin disks, $n_\perp \gg n_\parallel$, and when the aspect ratio (thickness to radius) of a disk tends to zero (thin film limit), we have $n_\perp \to 1, n_\parallel \to 0$. We find the following expression for susceptibility tensor $\hat{\chi}_{incl}$:

$$\hat{\chi}_{incl} = \begin{pmatrix} 0 & 0 & 0 \\ 0 & \dfrac{\omega_M (\omega_H + (n_\perp - n_\parallel)\omega_M)}{4\pi(\omega_0^2 - \omega^2)} & -\dfrac{i\omega_M \omega}{4\pi(\omega_0^2 - \omega^2)} \\ 0 & \dfrac{i\omega_M \omega}{4\pi(\omega_0^2 - \omega^2)} & \dfrac{\omega_H \omega_M}{4\pi(\omega_0^2 - \omega^2)} \end{pmatrix}$$

and using equation (9) we find the expression for permeability tensor $\hat{\mu}$ in the macrospin approximation:

$$\hat{\mu} = \begin{pmatrix} 1 & 0 & 0 \\ 0 & 1 + \dfrac{C}{\Omega_{\text{dyn}}^2 - \omega^2} & -\dfrac{i\rho\omega_M \omega}{\Omega_{\text{dyn}}^2 - \omega^2} \\ 0 & \dfrac{i\rho\omega_M \omega}{\Omega_{\text{dyn}}^2 - \omega^2} & 1 + \dfrac{B}{\Omega_{\text{dyn}}^2 - \omega^2} \end{pmatrix}, \quad (15)$$

where $\omega$ is the frequency of the incident electromagnetic wave, $C = \rho\omega_M \cdot [\omega_H + (n_\perp - n_\parallel + 2\xi)\omega_M]$, $B = \rho\omega_M \cdot [\omega_H - \xi\omega_M]$, $\omega_H = \gamma H_{\text{bias}}$, $\omega_M = \gamma 4\pi M$, and

$$\Omega_{\text{dyn}}^2 = \omega_0^2 - \xi\omega_M \left[(n_\perp - n_\parallel + 2\xi)\omega_M - \omega_H\right], \quad (16)$$

is the frequency of the uniform mode of the metamaterial calculated taking into account only the *internal dynamic* dipolar field produced by the lattice of magnetic nanoparticles, and

$$\omega_0^2 = \omega_H \left[\omega_H + (n_\perp - n_\parallel)\omega_M\right]$$

is the frequency of the uniform mode of an isolated inclusion (nanodisk). Using equation (16), we can find the frequency shift of the uniform mode of the metamaterial with respect to one of an isolated inclusion that is only due to the *internal dynamic* dipolar field produced by the lattice of magnetic inclusions:

$$\Delta\Omega_{\text{dyn}} \approx -\frac{\xi}{2} \cdot \frac{\omega_M \left[(n_\perp - n_\parallel + 2\xi)\omega_M - \omega_H\right]}{\omega_0}. \quad (17)$$

This shift is negative if $(n_\perp - n_\parallel + 2\xi)\omega_M > \omega_H$, which is usually true for magnetic fields smaller than $4\pi M$.

To take into account not only the dynamic but also static dipolar field produced by the other inclusions within the array, we should replace $\omega_{\text{dyn}}$ in equation (15) with the uniform resonance frequency of the metamaterial, $\Omega_0$, that is obtained from $\omega_{\text{dyn}}$ by substituting $\omega_H$ by $\omega_H + \xi\omega_M$ in equation (16). The result is given by

$$\Omega_0^2 = \omega_H \left(\omega_H + (n_\perp - n_\parallel + 3\xi)\omega_M\right), \quad (18)$$

while the total frequency shift relative to an isolated inclusion is

$$\Delta\Omega \approx \frac{3\xi}{2} \cdot \frac{\omega_M \omega_H}{\omega_0} \quad (19)$$

and is therefore positively defined, provided that both the dynamic and static internal dipolar fields within the array are taken into account.

### III. MICROMAGNETIC CALCULATION OF THE SUSCEPTIBILITIES OF AN ISOLATED INCLUSION AND A 2D ARRAY OF INCLUSIONS

The effective permeability of a metamaterial formed by a periodic array of magnetic inclusions can be obtained either directly from its susceptibility using equation (2), or from the susceptibility of an isolated inclusion using equation (9). From the point of view of micromagnetic calculations, the latter calculation is significantly simpler, takes less computational power, and can be performed using a greater variety of micromagnetic methods. The latter are well-developed, and we will show below that the results obtained with the *three* particular micromagnetic codes and with the dynamic matrix method agree for our case to within 2.7%.

There are two ways to calculate the susceptibility of the whole metamaterial using micromagnetic simulations. The first one is to simulate dynamic response of a sample that is large enough to be considered as effectively infinite. In this way, one could, in principle, obtain collective magnonic excitations with all possible values of the wave-vector. However, the method would require enormous computational power. The calculation of the effective permeability requires one to consider collective excitations formed by in-phase motion of meta-atoms, i.e. with the collective wave vector of zero. The response corresponding to such excitations can be calculated a different approach. Namely, we can perform micromagnetic simulations for a relatively small array to which periodic boundary conditions (PBCs) are applied in order to annihilate the influence of the array's edges. PBCs are realized differently in different micromagnetic packages (such as NMAG, MicroMagus, or OOMMF). However, the results obtained using the different packages have never been compared before. Hence, as will be discussed in the following, the susceptibility of the metamaterial obtained by this method is not as reliable as that of an isolated single inclusion.

### A. Simulations of a susceptibility of an isolated disk

To calculate the susceptibility tensor $\hat{\chi}_{incl}$ of an isolated inclusion, we have used two different methods. In the first of them, we have performed full-scale numerical micromagnetic calculations, in which the Landau-Lifshitz (LL) equation[7] is solved numerically in the time domain and the result is then Fourier transformed into the frequency domain. The second method is the dynamic matrix method (DMM), which in its initial form uses the diagonalization of the matrix computed from second derivatives of the system's energy, which is analogous to solving the linearized LLG equation[30]. In both methods, $\hat{\chi}_{incl}$ is calculated using its frequency domain definition

$$(\hat{\chi}_{incl})_{ij}(\omega) = \frac{\overline{m}_i(\omega)}{h'_j(\omega)}, \qquad (20)$$

where $\overline{m}_i(\omega)$ is the Fourier transform of the *i*-th component of the magnetization (spatially averaged over the volume of the inclusions) and $h'_j(\omega)$ is the Fourier transform of the *j*-th component of the external dynamic magnetic field.

The geometrical and magnetic parameters used in the calculations are listed in Table 1. Local magnetic field $\mathbf{h}'_0$ of 1 kOe is applied in the disk plane along the *x*-direction. From equation (14), the static local dipolar field for the filling factor of $\rho = 2.48\%$ is $h_{stat} = 67$ Oe. Hence, the calculation corresponds to the external magnetic bias field applied to the metamaterial $H_{bias} = 933$ Oe.

| The parameter | Value |
|---|---|
| disk diameter ($d$) | 195 nm |
| disk thickness ($l$) | 5 nm |
| saturation magnetization ($M$) | 800 G |
| exchange constant ($A$) | 1.3 $\mu$erg/cm |
| Gilbert damping constant ($\alpha$) | 0.01 |
| gyromagnetic ratio ($\gamma$) | $2\pi \cdot 2.8$ GHz/kOe |

**Table 1.** The geometrical and magnetic parameters used in the calculations are listed.

### A.1. Micromagnetic simulations

In simulations performed by all packages the disk shaped nanoelement was excited by a short transient magnetic field of the same temporal form, and the instantaneous spatially averaged magnetization was recorded every 10 ps over the time interval of 10.24 ns. However, the discretization cell size was different in simulations performed using different packages. In OOMMF, a uniform discretization into cells with dimensions of 1 x 1 x 5 nm$^3$ was used. In MicroMagus, the cell size was 2 x 2 x 5 nm$^3$, while the magnetization of the edge cells was additionally reduced to ensure a more adequate representation of the curved disk border. Due to the ability of the finite-element packages to use site-dependent meshes, in NMAG we could use an irregular mesh: small elements with the size of ~ 0.5 nm near the edges and larger elements with the typical sizes of ~ 2 nm at the center of the disk. It was also checked that further decrease of the cell size did not lead to significant changes in the calculated susceptibility.

To solve LLG equation, in MicroMagus we have used the Bogacki-Shampine version of the Runge-Kutta-23 method, which enables the integration step size control to achieve the required dynamical accuracy (in our case, we have set the accuracy to 10$^{-6}$ and checked that further accuracy improvement did not change the final result). In NMAG, we have used the second order BDF (backward differentiation formula) as implemented by the Sundials package[34]. The preconditioned Newton method is used to solve the implicit formula (these are the default settings in NMAG).

The excitation field **h'** was taken in the form of the 'sinc' function and applied along corresponding coordinate axes, in order to obtain various components of $\hat{\chi}_{\text{incl}}$:

$$\mathbf{h}' = h_{\max} \frac{\sin(2\pi f_0 (t-t_0))}{2\pi f_0 (t-t_0)}, \tag{21}$$

where amplitude $h_{\max}$ is 10 Oe, cut-off frequency $f_0$ is 30 GHz, and $t_0 = 10 / 2f_0$ ns. The pulse form given by equation (21) has a constant power spectrum up to cut-off frequency $f_0$. This feature assures that all system eigenmodes with frequencies $f < f_0$ are excited with an approximately constant 'strength'.

### A.2. Modified dynamical matrix method

The dynamical matrix method allows one to study spin wave modes of a magnetic particle of arbitrary shape and non-uniform equilibrium magnetization state, taking into account the external field, magnetic anisotropies, and the magneto-dipolar and exchange interactions[30]. In the present work, the original version of the method has been modified by including two additional terms: an external magnetic excitation at fixed frequency and the Gilbert damping. Thereby, the mathematical problem is changed from a generalized eigenvalue problem to a non-homogeneous linear system. Assuming that the magnetic particle is discretized into $N$ identical interacting cells, and that the uniform normalized magnetization in each cell $\mathbf{m}_i$ is represented by polar and azimuthal angles $\theta_i$ and $\varphi_i$, the equations of motion in the linear regime in the frequency domain become

$$\left( \hat{H} + \frac{M_S \omega}{\gamma} \hat{A} \right) \mathbf{v} = M_S \mathbf{w}, \quad (22)$$

where $\hat{H}$ is the Hessian matrix of the system, i.e. the matrix of second partial derivatives of the energy calculated in the ground state. Matrix $\hat{A}$ is given by

$$\hat{A} = \begin{pmatrix} i\alpha \sin^2 \varphi_1 & -i \sin \varphi_1 & 0 & 0 & \dots \\ i \sin \varphi_1 & i\alpha & 0 & 0 & \dots \\ 0 & 0 & i\alpha \sin^2 \varphi_2 & -i \sin \varphi_2 & \dots \\ 0 & 0 & i \sin \varphi_2 & i\alpha & \dots \\ \dots & \dots & \dots & \dots & \dots \end{pmatrix},$$

$\mathbf{v}$ is the vector of the magnetization fluctuations (variables)

$$\mathbf{v} = \begin{pmatrix} \delta\theta_1 \\ \delta\varphi_1 \\ \vdots \\ \delta\theta_N \\ \delta\varphi_N \end{pmatrix},$$

and vector $\mathbf{w}$ in the right-hand side is

$$\mathbf{w} = \begin{pmatrix} \mathbf{h}' \cdot \frac{\partial \mathbf{m}_1}{\partial \theta_1} \\ \mathbf{h}' \cdot \frac{\partial \mathbf{m}_1}{\partial \varphi_1} \\ \vdots \\ \mathbf{h}' \cdot \frac{\partial \mathbf{m}_1}{\partial \theta_N} \\ \mathbf{h}' \cdot \frac{\partial \mathbf{m}_1}{\partial \varphi_N} \end{pmatrix},$$

The new terms with respect to the original formulation (Ref. 30) are non-homogeneous term $\mathbf{w}$ and the diagonal elements in $\hat{A}$. Here, $\mathbf{h}'$ is the homogeneous external field oscillating at frequency $\omega$ as discussed above. Linear system of equations (22) can be solved for any frequency, and corresponding average magnetization $\overline{\mathbf{m}} = M_S \sum_i \mathbf{m}_i / N$ is then used to find the susceptibility defined by equation (20).

For the numerical solution of equation (22), we used an iterative method. Finding the solution of a large, complex, non-hermitian system of linear algebraic equations is a difficult task, especially close to singular points, corresponding to frequencies of self-oscillations of the magnetic particle. An iterative method allows one to use the result obtained at a given frequency as the initial value for the search of the solution at the next frequency, thereby reducing the convergence time. We have chosen the bi-conjugate gradient squared method with stabilization[35]. This method refers the matrix of the system only through its multiplication by a vector, or the multiplication of its transpose conjugate by a vector, so that we can easily exploit the symmetries of matrices $\hat{H}$ (real, symmetric) and $\hat{A}$ (sparse, tridiagonal) for an efficient implementation of the algorithm.

The susceptibility of the isolated disk shaped nanoelement has been calculated using a uniform discretization into cells with dimensions of 3 x 3 x 5 nm$^3$. Albeit larger than that used for the other methods, this cell size allows us to reduce the number of independent variables in the numerical problem to a manageable size. In fact, despite the unquestionable advantages of the iterative method chosen for solving the complex system of equations, its convergence is not guaranteed, in particular failing when the number of variables is too large.

### B. Micromagnetic simulations using periodic boundary conditions

To calculate the susceptibility tensor $\hat{\chi}$ from simulations with PBCs, we have used two micromagnetic codes: NMAG and MicroMagus. We have performed the simulations with the unit cell of two magnetic disks and periodic boundary conditions. PBCs are realised differently in different micromagnetic packages. The underlying aim is, however, always the same: to reduce the effect of the finite size, i.e. by modifying the internal field in a finite array so that the array can be treated as a part of an infinite one.

In NMAG, PBCs are realized through the possibility to create a finite number of virtual copies of the simulated object[36] (two disks in our case). The magnetization dynamics in every virtual copy repeats the one in the simulated object. The magnetization dynamics in the simulated object is simulated taking into account both static and dynamic dipolar fields produced by the virtual copies. In the NMAG simulations reported here, we created as many virtual copies as to obtain a virtual array of 7x5 disks, which corresponds to the virtual sample size of ~1.5 x1.7 $\mu m^2$. Furthermore, we have checked that a further increase of the size of the virtual array to 11x7 or ~2.3 x2.4 $\mu m^2$ leads only to an insignificant increase of the frequency by 0.016 GHz.

In MicroMagus, PBCs are taken into account using the rigorous Ewald method for the magneto-dipolar interaction, developed by the package authors initially for a 2D lattice of point dipoles[37] and extended later to the case of a system discretization by finite rectangular prisms[38]. For this reason, the simulation area used by MicroMagus includes only one elementary lattice cell of two nanodiscs.

The susceptibility $\hat{\chi}$ is calculated using an equation similar to Eq. (20):

$$\hat{\chi}_{ij}(\Omega) = \frac{\overline{m}_i(\Omega)}{h'_j(\Omega)}, \qquad (23)$$

with the only difference that the frequency $\Omega$ here is the metamaterial excitation frequency, while the excitation frequency of a single inclusion $\omega$ is used in Eq. (20).

## IV. RESULTS AND DISCUSSION

In Fig. 2, four components of the permeability tensor of the metamaterial calculated using equation (9) are presented. The blue solid lines, dashed red lines, green open dots, and black filled dots show the permeability calculated using OOMMF, NMAG, MicroMagus, and DMM, respectively. For the in-plane edge-to-edge separation $b = 20$ nm assumed in the calculations and the interlayer distance $c = 140$ nm, the filling factor of the magnetic material is $\rho = 2.48$ %. One can clearly see the contributions from the two main resonances of the nanodisk (meta-atoms) into the effective permeability tensor of the metamaterial. The first (lower frequency) resonance corresponds to the edge mode[39], for which the magnetization precesses mainly near the edges and the precession amplitude is nearly zero in the central part of the disk. This mode does not have nodal lines. The second resonance corresponds to the first bulk mode with two nodal lines perpendicular to the direction of the bias field, which is directed horizontally in Fig. 2. The mode profiles[40] are shown in the insets in Fig. 2. The inset also shows the distribution of the normalized magnetization component orthogonal to the applied field (which is horizontal in our case) in the ground state in the single nanodisk. This picture shows that the applied field nearly saturates the sample.

The bias magnetic field of $H_{\text{bias}} = 933$ Oe is applied in the plane of the metamaterial layers along the $x$ axis. This field is sufficiently large to achieve an almost saturated equilibrium magnetization state of the sample. For this reason, the real part of the $xx$ component of the effective permeability differs insignificantly from unity and the $xy$, $yx$, $xz$, and $zx$ components are almost zero, with the latter differences being due to small ground state nonuniformities caused by the strong demagnetizing field near the edges of metamaterial constituents. Hence, these components are not displayed in Fig. 2. The $yy$ and $zz$ components differ in magnitude by more than 7 times. This difference is due to the ellipticity of the precession (caused by the demagnetizing field of the thin nanodisc), with the in-plane oscillation amplitude being much larger than that of the out-of-plane oscillation. As shown in Fig. 2, the $yy$ component of the effective permeability becomes negative near both resonant frequencies at the particular value of the filling factor.

As one can see from Fig. 2, all three micromagnetic packages and the DMM yield very similar results for the frequency dependence of the permeability tensor components. To compare the results given by different methods quantitatively way, Table 2 summarizes the frequencies of the permeability resonances corresponding to the two dominant modes of the corresponding isolated nanodisc ("meta-atom"). For the edge mode (the lowest frequency mode), the greatest discrepancy is observed between predictions of OOMMF and Micromagus, with the difference of 0.2 GHz or about 2.7% of the average edge mode frequency. The difference is most likely to result from the different discretization methods of the disk edges used in the calculations by the different packages. Indeed, the edge mode is strongly localized near the disk edges, where the internal dipolar field varies significantly over distances of a couple of nanometers from the edges. So, the discretization method and the treatment of the dipolar field created by the edge cells play an important role for the edge mode[41]. In particular, we found that for the studied sample the frequency of the edge mode decreased in NMAG, increased in OOMMF, and remained nearly constant in MicroMagus as the cell size of the underlying mesh decreased. In NMAG and OOMMF, we have checked that a decrease of the cell size beyond that used in the calculations reported here does not lead to significant changes in the calculated resonant frequencies.

For the bulk mode (the higher frequency mode), all three micromagnetic packages give identical predictions. This result is expected, because the mode occupies predominantly the inner region of the disk, so that the details of the edge discretization play a minor role. The DMM predicts a value that is by 0.23 GHz (or about 2.3 % of the average bulk mode frequency) larger than that predicted by the micromagnetic packages. The difference is probably due to the larger cell size used in the DMM calculations, which makes the representation of the mode with oscillations less accurate. The iterative algorithm used in this case for solving the general, complex, linear system involved in the DMM may also have a role, introducing numerical errors (for this algorithm an error around one percent is quite possible). More information on the influence of the mesh sizes on different modes in confined magnetic elements can be found, e.g., in Ref. 41.

| Package | Edge mode | Bulk mode |
| --- | --- | --- |
| OOMMF | 7.35 GHz | 9.82 GHz |
| NMAG | 7.49 GHz | 9.83 GHz |
| MicroMagus | 7.55 GHz | 9.82 GHz |
| DMM | 7.46 GHz | 10.05 GHz |

**Table 2.** The resonant frequencies of the two dominant modes of an isolated nanodisc obtained by different micromagnetic packages and the DMM are shown.

The resonant frequencies in the permeability of the metamaterial are shifted by about $\Delta\Omega \approx 0.17\,\text{GHz}$ relative to those of the corresponding spin wave modes of an isolated nanodisk. As discussed for the macrospin calculations, the shift contains two competing contributions of the dynamic and static dipolar fields. The contribution from the dynamic fields is negative and has been estimated by a straight application of equation (9) to the susceptibility of the nanodisk, as about $\Delta\Omega_{\text{dyn}} \approx -0.12\,\text{GHz}$, varying insignificantly for the different modes and packages. The effect of the increase of the static internal field by 67 Oe has been estimated by calculating the mode frequencies for an isolated nanodisc at $H_{\text{bias}}$ = 933 Oe and 1 kOe, to result in an increase of the frequency by about $\Delta\Omega_{\text{stat}} = 0.29\,\text{GHz}$. So, dynamic and static dipolar fields within the array lead to frequency shifts of opposite signs, and together they lead to the observed net increase of the resonant frequencies of the metamaterial of about 0.17 GHz.

Let us compare the results of the full micromagnetic calculations and those obtained using the macrospin approximation. The frequency of the only resonance predicted for the metamaterial by the macrospin model is calculated using equation (18) and the parameter values given in Table 1 to yield $\Omega_0 = 9.02\,\text{GHz}$. The frequency is about 1 GHz lower and about 1.5 GHz higher than those of the bulk and edge modes in the micromagnetic calculations respectively. Furthermore, the macrospin model predicts a total shift (equation (19)) of $\Delta\Omega \approx 0.08\,\text{GHz}$, which includes a contribution due to the dynamic dipolar field (equation (17)) of $\Delta\Omega_{dyn} \approx -0.27\,\text{GHz}$. Perhaps expectedly, the results demonstrate the failure of the macrospin approximation to predict the microwave permeability and more generally dynamic magnetic properties of magnonic metamaterials composed of non-ellipsoidal magnetic inclusions. In contrast, the proposed method based on full micromagnetic calculations allows us to take into account contributions of all eigenmodes of constituent magnetic inclusions of a non-ellipsoidal shape to the effective permeability. Yet, the semi-analytical nature of the micromagnetic model allows us to compare the contributions of dynamic and static dipolar fields into the observed frequency dependence of the permeability.

The results of micromagnetic simulations performed using periodic boundary conditions (PBCs) are summarized in Fig. 3, in which we present the zz component of the effective permeability tensor calculated using equation (2) with the susceptibility tensor calculated from equation (23). The red dashed and blue dashed lines show the permeability of the metamaterial consisting of non-interacting magnetic disks based on the susceptibility of a single disk calculated by NMAG and MicroMagus, respectively. Red solid and black solid lines show the permeability of the metamaterial calculated by NMAG and MicroMagus respectively using PBCs as explained in sec. III.B. Thereby Fig. 3 demonstrates the effect of the interaction between metamaterial constituents on resonant frequencies of the metamaterial. Table 3 shows the frequency shifts of the dominant modes of the metamaterial with respect to the frequencies of the corresponding modes of an isolated nanodisc. The frequency shifts predicted by the two packages differ by 0.04 GHz for the edge mode and 0.1 GHz for the bulk mode. The difference is insignificant given the discrepancy of 0.06 GHz and 0.01 GHz in predictions of the same two packages for the edge and bulk modes of an isolated disk (open boundary conditions) respectively. Hence, the difference in the edge mode frequency between these two packages in simulations with PBC can be still attributed to the same reason as for an isolated disk, i.e. to the different methods of the edge discretization. The relatively large difference between the bulk mode frequencies in simulations with PBC performed by NMAG and MicroMagus can be only due to different algorithms of the PBC realization in these packages.

| Package | The shift of the edge mode | The shift of the bulk mode |
| --- | --- | --- |
| NMAG | 0.91 GHz | 0.3 GHz |
| MicroMagus | 0.87 GHz | 0.4 GHz |

**Table 3.** The frequency shifts for both dominant modes of the metamaterial with respect to those of an isolated nanodisc are shown for simulations with PBCs using NMAG and MicroMagus.

The shift of the lower metamaterial resonant frequency with respect to the edge mode of an isolated nanodisc predicted by the micromagnetic simulations with PBCs is significantly larger than that in the proposed semi-analytical model. At the same time, the micromagnetic simulations with PBC predict a shift of the higher metamaterial resonant frequency with respect to the bulk mode of a nanodisc that is again larger than and yet comparable with that in the semi-analytical model. This can be explained by the fact that the bulk mode is localized in the inner part of the disk where the dipolar field produced by the other discs in the metamaterial is described reasonably well by the dipolar approximation used in the calculations.

When studying metamaterials, frequency domains in which negative values of the effective permeability can be achieved are especially important, due to the possibility of realizing negative refraction. The diagram in Fig. 4 shows the range of values of the in-plane edge-to-edge separation (b) and the distance between layers (c) in which the $\mu_{yy}$ component becomes negative in the vicinity of the both resonances. In particular, the diagram demonstrates that, in order to provide negative $\mu_{yy}$ at the corresponding frequencies, the filling factor should be greater than $\rho_{\min} = 1.6\%$.

## V. CONCLUSIONS

We have shown that full micromagnetic simulations and calculations based on the dynamical matrix method provide a useful complementary tool to the well-developed averaging procedures widely used in the theory of the effective permeability. For the model case of an array of ferromagnetic discs in a non-magnetic matrix, we have been able to calculate the effective permeability tensor taking into account all resonances of the metamaterial constituents. We have compared the proposed suggested model with that based on the macrospin approximation and with the direct permeability calculation based on micromagnetic simulations with periodic boundary conditions. The region of geometric parameters where the effective permeability of the studied metamaterial can reach negative values has also been determined. In addition, the calculation provides a useful method by which to evaluate the accuracy of the different micromagnetic packages and the dynamical matrix method. In particular, we find that the results produced by the state-of-art micromagnetic simulations agree with each other within an error bar of about 3 %, which has to be taken into account when micromagnetic calculations are used to model experimental data.

The research leading to these results has received funding from the ECs 7th Framework Programme (FP7/2007-2013) under Grant Agreements 228673 (MAGNONICS) and 233552 (DYNAMAG).

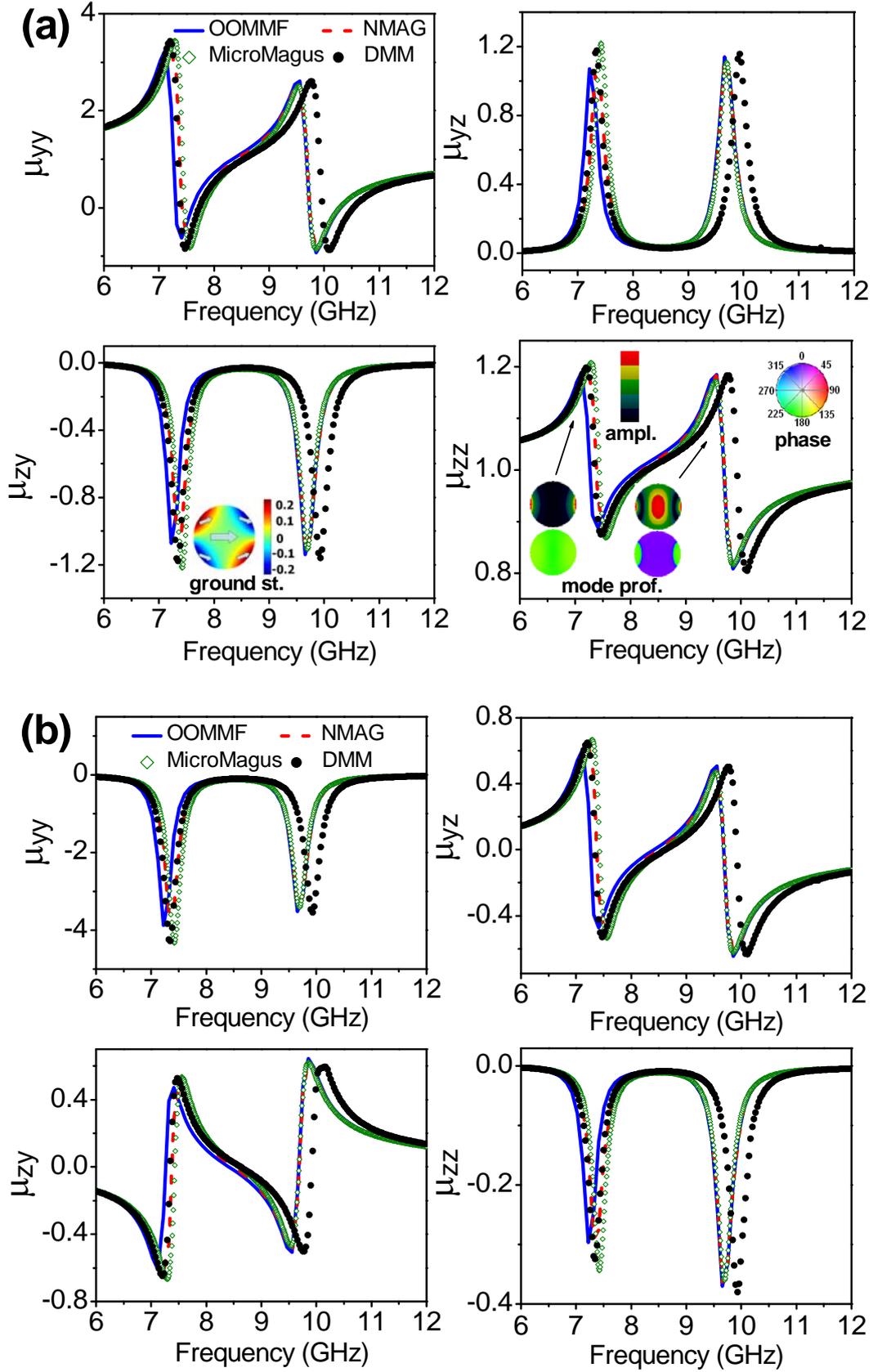

**Fig. 2.** (Color online) The real **(a)** and imaginary **(b)** parts of four components of effective permeability tensor $\mu$ calculated using equation (9) are shown as functions of the frequency for the metamaterial depicted in Fig. 1. External magnetic field $H_{\text{bias}} = 933$ Oe is applied in the plane of the layers along

horizontal line. The blue solid lines, dashed red lines, green open dots, and black filled dots show the permeability calculated using OOMMF, NMAG, MicroMagus, and DMM, respectively. The insets show the distribution of the normalized magnetization component orthogonal to the applied field in the ground state in a single nanodisk; the spatial profiles of the mode amplitude (top) and phase (bottom) for the two dominant modes of a single nanodisk.

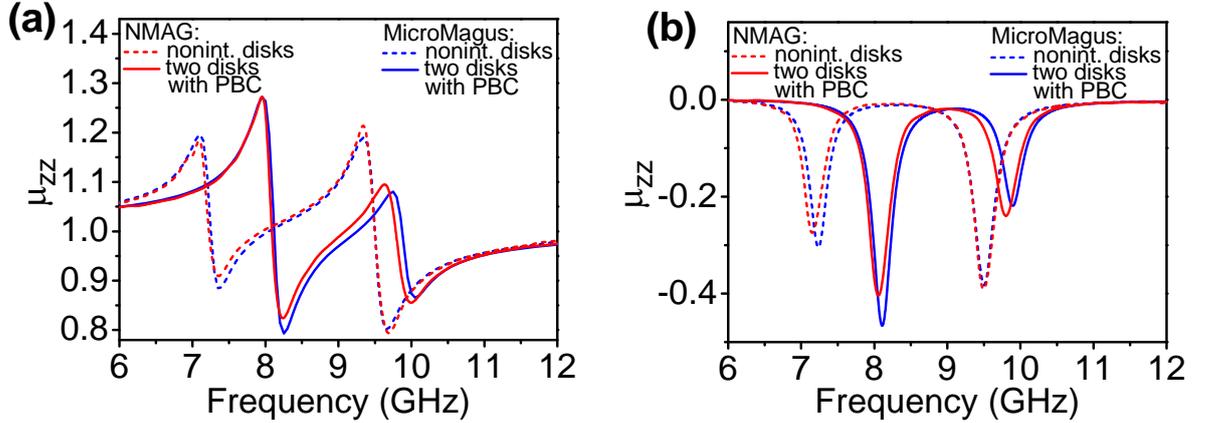

**Fig. 3.** (Color online) The real (**a**) and imaginary (**b**) parts of the *zz* component of effective permeability tensor $\mu$ calculated using equation (2) with the susceptibility tensor calculated using equation (23) are shown as functions of the frequency for the metamaterial depicted on Fig. 1. External magnetic field $H_{bias}$ = 933 Oe is applied in the plane of the layers parallel to the *x* axis. The red dashed and blue dashed lines show the permeability of the metamaterial consisting of non-interacting magnetic disks based on the susceptibility of a single disk calculated by NMAG and MicroMagus respectively. The red solid and black solid lines show the permeability of the metamaterial calculated using periodic boundary conditions by NMAG and MicroMagus respectively.

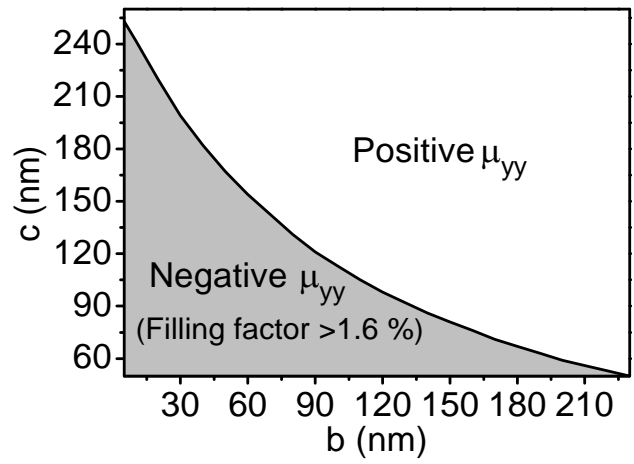

**Fig. 4.** The range of values of the in-plane edge-to-edge separation (b) and the distance between layers (c) in which $\mu_{yy}$ component becomes negative in the vicinity of the both resonances.

1       V. G. Veselago, Sov. Phys. Uspekhi **10**, 509 (1968).

2       J. B. Pendry, D. Schurig, and D. R. Smith, Science **312**, 1780 (2006).

3       J. B. Pendry, Phys. Rev. Lett. **85**, 3966 (2000).

4       J. B. Pendry, A. J. Holden, D. J. Robbins, and W. J. Stewart, IEEE Trans. Microwave Theory Tech. **47**, 2075 (1999).

5       O. Acher, J. Magn. Magn. Mater. **321**, 2093 (2009).

6       R. V. Mikhaylovskiy, E. Hendry, and V. V. Kruglyak, Phys. Rev. B **82**, 195446 (2010).

7       A. G. Gurevich and G. A. Melkov, Magnetization oscillations and waves (CRC Press, 1996).

8       S. V. Vasiliev, V. V. Kruglyak, M. L. Sokolovskii, and A. N. Kuchko, J. Appl. Phys. **101**, 113919 (2007).

9       V. V. Kruglyak, S. O. Demokritov, and D. Grundler, J. Phys. D: Appl. Phys. **43**, 264001 (2010).

10      U. Ebels, J. L. Duvail, P. E. Wigen, L. Piraux, L. D. Buda, and K. Ounadjela, Phys. Rev. B **64**, 144421 (2001).

11      G. S. Makeeva, M. Pardavi-Horvath, and O. A. Golovanov, IEEE Trans. Magn. **45**, 4074 (2009).

12      V. Boucher, L. – P. Carignan, T. Kodera, C. Caloz, A. Yelon, and D. Ménard, Phys. Rev. B **80**, 224402 (2009).

13      S. Rajagopalan and J. K. Furdyna, Phys. Rev. B **39**, 2532 (1989).

14      A. Aharoni, J. Appl. Phys. **69**, 7762 (1991); *ibid*. **81**, 830 (1997).

15      G. Viau, F. Fievet-Vincent, F. Fievet, P. Toneguzzo, F. Ravel, and O. Acher, J. Appl. Phys. **81**, 2749 (1997).

16      D. Mercier, J. - C. S. Levy, G. Viau, F. Fievet-Vincent, F. Fievet, P. Toneguzzo, and O. Acher, Phys. Rev. B **62**, 532 (2000).

17      J. Ramprecht and D. Sjöberg, J. Phys. D – Appl. Phys. **41**, 135005 (2008).

18      R. D. McMichael and M. D. Stiles, J. Appl. Phys. **97,** 10J901 (2005).

19      V. V. Kruglyak, P. S. Keatley, A. Neudert, M. Delchini, R. J. Hicken, J. R. Childress, and J. A. Katine, Phys. Rev. B **77,** 172407 (2008).

20      F. Montoncello, L. Giovannini, F. Nizzoli, H. Tanigawa, T. Ono, G. Gubbiotti, M. Madami, S. Tacchi, and G. Carlotti, Phys. Rev. B **78**, 104421 (2008).

21      V. V. Kruglyak, P. S. Keatley, A. Neudert, R. J. Hicken, J. R. Childress, and J. A. Katine, Phys. Rev. Lett. **104**, 027201 (2010).

22      J. Topp, D. Heitmann, M. P. Kostylev, and D. Grundler, Phys. Rev. Lett. **104**, 207205 (2010).

[23] E. J. Kim, J. L. R. Watts, B. Harteneck, A. Scholl, A. Young, A. Doran, and Y. Suzuki, J. Appl. Phys. **109**, 07D712 (2011).

[24] V. L. Zhang, Z. K. Wang, H. S. Lim, S. C. Ng, M. H. Kuok, S. Jain, and A. O. Adeyeye, J. Nanosci. Nanotechn. **11**, 2657 (2011).

[25] P. Rovillain, R. de Sousa, Y. Gallais, A. Sacuto, M. A. Measson, D. Colson, A. Forget, M. Bibes, M. Barthelemy, and M. Cazayous, Nature Mater. **9**, 975 (2010).

[26] A. Kumar, J. F. Scott, and R. S. Katiyar, Appl. Phys. Lett. **99**, 062504 (2011).

[27] T. Fischbacher, M. Franchin, G. Bordignon, and H. Fangohr. IEEE Trans. Magn. **43**, 2896 (2007), http://NMAG.soton.ac.uk.

[28] M. Donahue, and D.G. Porter, OOMMF User's guide, Version 1.0, Interagency Report NISTIR 6376, NIST, 1999, http://math.nist.gov/oommf.

[29] D. V. Berkov, and N. L. Gorn, MicroMagus – package for micromagnetic simulations, http://www.micromagus.de.

[30] M. Grimsditch, L. Giovannini, F. Montoncello, F. Nizzoli, G. K. Leaf, and H. G. Kaper, Phys Rev. B **70**, 054409 (2004).

[31] L. Giovannini, F. Montoncello, and F. Nizzoli, Phys Rev. B **75**, 024416 (2007).

[32] K. Rivkin, W. Saslow, L. E. De Long, and J. B. Ketterson, Phys Rev. B **75**, 174408 (2007).

[33] M. Dvornik, P. V. Bondarenko, B. A. Ivanov, and V. V. Kruglyak, J. Appl. Phys. **109**, 07B912 (2011).

[34] https://computation.llnl.gov/casc/sundials/documentation/cv_guide.pdf

[35] H. A. van der Vorst, SIAM J. Sci. Statist. Comput. **13**, 631 (1992).

[36] H. Fangohr, G. Bordignon, M. Franchin, A. Knittel, P. A. J. de Groot, and T. Fischbacher, J. Appl. Phys. **105**, 07D529 (2009).

[37] D. V. Berkov, N. L. Gorn, Phys. Rev. B **57**, 14332 (1998)

[38] D. V. Berkov, N. L. Gorn, IEEE Trans. Magn. **MAG-38,** 2474 (2002)

[39] V. V. Kruglyak, A. Barman, R. J. Hicken, J. R. Childress, and J. A. Katine, Phys. Rev. B **71,** 220409 (2005).

[40] The mode profiles were calculated using SEMARGL (http://www.magnonics.org/semargl/).

[41] R. E. Camley, B. V. McGrath, Y. Khivintsev, Z. Celinski, R. Adam, C. M. Schneider, and M. Grimsditch, Phys. Rev. B **78**, 024425 (2008).